\documentclass[%
 reprint,
superscriptaddress,
 amsmath,amssymb,
 aps,
prc,
]{revtex4-1}

\usepackage{graphicx}
\usepackage{dcolumn}
\usepackage{bm}

\usepackage{booktabs}
\newcolumntype{C}[1]{>{\centering\arraybackslash}p{#1}}
\AtBeginDocument{
\heavyrulewidth=.08em
\lightrulewidth=.05em
\cmidrulewidth=.03em
\belowrulesep=.65ex
\belowbottomsep=0pt
\aboverulesep=.4ex
\abovetopsep=0pt
\cmidrulesep=\doublerulesep
\cmidrulekern=.5em
\defaultaddspace=.5em
}

\begin{document}

\preprint{APS/123-QED}

\title{Study of the $\beta$-decay of $^{100}$Tc with Total Absorption $\gamma$-Ray Spectroscopy}

\author{V.~Guadilla}
\email{guadilla@ific.uv.es}
\affiliation{%
 Instituto de F\'isica Corpuscular, CSIC-Universidad de Valencia, E-46071, Valencia, Spain
}
\author{A.~Algora}%
\email{algora@ific.uv.es}
\affiliation{%
 Instituto de F\'isica Corpuscular, CSIC-Universidad de Valencia, E-46071, Valencia, Spain
}
\affiliation{%
 Institute of Nuclear Research of the Hungarian Academy of Sciences, Debrecen H-4026, Hungary.
}
\author{J.L.~Tain}%
\author{J.~Agramunt}  
\author{D.~Jordan} 
\author{A.~Montaner-Piz\'a}  
\author{S.E.A.~Orrigo}  
\author{B.~Rubio}  
\author{E.~Valencia}  
\affiliation{%
 Instituto de F\'isica Corpuscular, CSIC-Universidad de Valencia, E-46071, Valencia, Spain
}
\author{J.~Suhonen}
\affiliation{%
 University of Jyvaskyla, Department of Physics, P.O. Box 35, FI-40014 University of Jyvaskyla, Finland
}
\author{O.~Civitarese}
\affiliation{%
 Department of Physics, University of La Plata, C.C. 67 1900, La Plata, Argentina
} 
\author{J.~\"Ayst\"o}  
\affiliation{%
 University of Jyvaskyla, Department of Physics, P.O. Box 35, FI-40014 University of Jyvaskyla, Finland
}
\author{J.A.~Briz}  
\affiliation{%
 Subatech, CNRS/IN2P3, Nantes, EMN, F-44307, Nantes, France
}
\author{A.~Cucoanes}  
\affiliation{%
 Subatech, CNRS/IN2P3, Nantes, EMN, F-44307, Nantes, France
}
\author{T.~Eronen}  
\affiliation{%
 University of Jyvaskyla, Department of Physics, P.O. Box 35, FI-40014 University of Jyvaskyla, Finland
}
\author{M.~Estienne}  
\affiliation{%
 Subatech, CNRS/IN2P3, Nantes, EMN, F-44307, Nantes, France
}
\author{M.~Fallot}  
\affiliation{%
 Subatech, CNRS/IN2P3, Nantes, EMN, F-44307, Nantes, France
}
\author{L.M.~Fraile}  
\affiliation{%
Universidad Complutense, Grupo de F\'isica Nuclear, CEI Moncloa, E-28040, Madrid, Spain
}
\author{E.~Ganio\u{g}lu}  
\affiliation{%
Department of Physics, Istanbul University, 34134, Istanbul, Turkey
}
\author{W.~Gelletly}  
\affiliation{%
 Instituto de F\'isica Corpuscular, CSIC-Universidad de Valencia, E-46071, Valencia, Spain
}
\affiliation{%
Department of Physics, University of Surrey, GU2 7XH, Guildford, UK
} 
\author{D.~Gorelov}  
\author{J.~Hakala} 
\author{A.~Jokinen}  
\author{A.~Kankainen}  
\author{V.~Kolhinen}  
\author{J.~Koponen}  
\affiliation{%
 University of Jyvaskyla, Department of Physics, P.O. Box 35, FI-40014 University of Jyvaskyla, Finland
}
\author{M.~Lebois}  
\affiliation{%
Institut de Physique Nucl\`eaire d'Orsay, 91406, Orsay, France
}
\author{T.~Martinez}  
\affiliation{%
Centro de Investigaciones Energ\'eticas Medioambientales y Tecnol\'ogicas, E-28040, Madrid, Spain
}
\author{M.~Monserrate}  
\affiliation{%
 Instituto de F\'isica Corpuscular, CSIC-Universidad de Valencia, E-46071, Valencia, Spain
}
\author{I.~Moore}  
\affiliation{%
 University of Jyvaskyla, Department of Physics, P.O. Box 35, FI-40014 University of Jyvaskyla, Finland
}
\author{E.~N\'acher}  
\affiliation{%
Instituto de Estructura de la Materia, CSIC, E-28006, Madrid, Spain
}
\author{H.~Penttil\"a}  
\author{I.~Pohjalainen}  
\affiliation{%
 University of Jyvaskyla, Department of Physics, P.O. Box 35, FI-40014 University of Jyvaskyla, Finland
}
\author{A.~Porta}  
\affiliation{%
 Subatech, CNRS/IN2P3, Nantes, EMN, F-44307, Nantes, France
}
\author{J.~Reinikainen}  
\author{M.~Reponen}  
\author{S.~Rinta-Antila}  
\author{K.~Rytk\"onen}  
\affiliation{%
 University of Jyvaskyla, Department of Physics, P.O. Box 35, FI-40014 University of Jyvaskyla, Finland
}
\author{T.~Shiba}  
\affiliation{%
 Subatech, CNRS/IN2P3, Nantes, EMN, F-44307, Nantes, France
}
\author{V.~Sonnenschein}  
\affiliation{%
 University of Jyvaskyla, Department of Physics, P.O. Box 35, FI-40014 University of Jyvaskyla, Finland
}
\author{A.A.~Sonzogni}  
\affiliation{%
NNDC, Brookhaven National Laboratory, Upton, NY 11973-5000, USA
}
\author{V.~Vedia}  
\affiliation{%
Universidad Complutense, Grupo de F\'isica Nuclear, CEI Moncloa, E-28040, Madrid, Spain
}
\author{A.~Voss} 
\affiliation{%
 University of Jyvaskyla, Department of Physics, P.O. Box 35, FI-40014 University of Jyvaskyla, Finland
}
\author{J.N.~Wilson}
\affiliation{%
Institut de Physique Nucl\`eaire d'Orsay, 91406, Orsay, France
}
\author{A.-A.~Zakari-Issoufou} 
\affiliation{%
 Subatech, CNRS/IN2P3, Nantes, EMN, F-44307, Nantes, France
}

\date{\today}

\begin{abstract}
The $\beta$-decay of $^{100}$Tc has been studied using the Total Absorption  $\gamma$-Ray Spectroscopy technique at IGISOL. In this work the new DTAS spectrometer in coincidence with a cylindrical plastic $\beta$ detector has been employed. The $\beta$-intensity to the ground state obtained from the analysis is in good agreement with previous high-resolution measurements. However, differences in the feeding to the first excited state as well as weak feeding to a new level at high excitation energy have been deduced from this experiment. Theoretical calculations performed in the quasiparticle random-phase approximation (QRPA) framework are also reported. Comparison of these calculations with our measurement serves as a benchmark for calculations of the double $\beta$-decay of $^{100}$Mo.
\end{abstract}

\maketitle


\section{Introduction}

\subsection{Motivation}

The study of double $\beta$-decay processes is an interesting and challenging topic in nuclear and particle physics. It is amongst the rarest forms of radioactive decay and its occurrence has significant implications for the Standard Model of fundamental interactions. Double $\beta$-decay is a radioactive decay process in which a nucleus with proton and neutron numbers $(Z, N)$ undergoes a transition to the nucleus with $(Z+2, N-2)$. It can be observed for some nuclei, such as $^{100}$Mo, where the nucleus with atomic number higher by one unit ($^{100}$Tc) has a smaller binding energy, and the single $\beta$-decay is forbidden. If the nucleus with atomic number higher by two units, $^{100}$Ru, has a larger binding energy, then the double $\beta$-decay process is allowed energetically \cite{REPORT_TEO} (and references therein). 

With the exception of one unconfirmed case \cite{Klapdor}, double $\beta$-decay has so far only been detected in the so-called two-neutrino mode, when two electron antineutrinos are emitted in addition to the two electrons. This process occurs whether or not neutrinos are their own antiparticles (whether or not they are Majorana particles). On the other hand, the neutrinoless case of the decay, that would violate lepton-number conservation, is considered one of the best candidates to provide information about the absolute neutrino mass scale and the Dirac or Majorana nature of the neutrino (see \cite{Avignone2008,Vergados2012} for recent experimental and theoretical accounts of the subject). To extract this information one would need to determine experimentally the half-life of this very slow decay and estimate theoretically the phase-space factors and nuclear matrix elements (NME) implicit in the process. 

Theoretical calculations of the NME for double $\beta$-decay have been performed in the past using several approaches: the quasiparticle-random-phase-approximation (QRPA), the interacting shell model (ISM), the proton-neutron interacting boson model (IBA-2), the energy density functional approach (EDF) and the projected Hartree-Fock-Bogoliubov (HFB) mean-field scheme. A recent comparison of the different results can be found in the review \cite{Suhonen2012b}. 

The calculations of the NME require a knowledge of the wave functions of the nuclear states involved. 
It has been suggested that it is possible to test the accuracy of the neutrinoless (0$\nu$) double $\beta$-decay calculations by comparing the two-neutrino (2$\nu$) double $\beta$-decay calculations (within the same theoretical frameworks) with measured 2$\nu$ double $\beta$-decay rates. In the QRPA calculations the parameters of the model can be determined not only by using the double $\beta$-decay rates, but also using information on the single $\beta$-decay rates ($\beta^+$/EC, and $\beta^{-}$) for the intermediate nucleus. Precise data for the single $\beta$ decay of the associated intermediate nuclei of the double $\beta$-decay process can help to fix the effective value of the axial vector coupling constant, $g_{A}$, together with the value of the particle-particle strength, $g_{pp}$, within a QRPA framework. For this reason, improving our experimental knowledge of $\beta$-decays, both double and single $\beta$ decays that are relevant to the neutrinoless double $\beta$-decay calculations, should be considered to be of high priority. This is the main goal of the present work: an improved study of the single $\beta$-decay $^{100}$Tc $\rightarrow$ $^{100}$Ru to provide experimental constraints on nuclear-model calculations of the double $\beta$-decay of $^{100}$Mo. 

It should also be noted that constraining the parameters of the calculations is not just possible 
by means of $\beta$-decay studies. In recent years a great effort has also been invested in studies of the properties of ground-state wave functions of double $\beta$-decay candidates. For example, the occupancies of valence single-particle orbitals and pairing correlations of the states of interest have been measured by means of transfer reactions (see for example \cite{Freeman2012,Thomas2012}).

\begin{figure}[h]
\includegraphics[width=0.5 \textwidth]{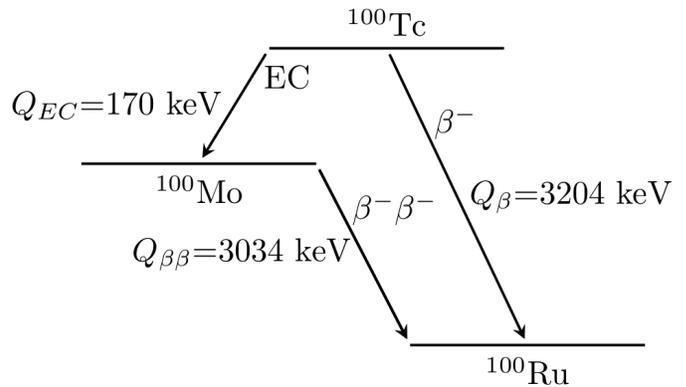}
\caption{\label{100Scheme} Schematic picture of the A=100 double $\beta$ decay system. The $Q_{\beta/EC}$ values are taken from \cite{Qval_NNDC}.}
\end{figure}

One of the best known double $\beta$-decay systems is the A=100 system shown in Figure \ref{100Scheme} ($^{100}$Mo, $^{100}$Tc, $^{100}$Ru) \cite{100TEO_1,100TEO_2,100TEO_3,100TEO_4,100TEO_5}. Double $\beta$-decay rates to the ground state and to the first excited $0^+$ state of $^{100}$Ru \cite{DB100} have been measured in the NEMO 3 experiment \cite{NEMO3}. The EC branch of the decay $^{100}$Tc $\rightarrow$ $^{100}$Mo has been measured recently with much higher precision than before \cite{EC100}. Also in a recent measurement using charge-exchange reactions the $^{100}$Ru $\rightarrow$ $^{100}$Tc transitions have been measured, indicating the nature of the single-state dominance in the double $\beta$-decay process \cite{SinStateDom_DB_1,100TEO_3}. On the other hand, the decay of $^{100}$Tc $\rightarrow$ $^{100}$Ru has only been measured using the high-resolution $\gamma$-ray spectroscopy technique \cite{100Tc_HR_1,100Tc_HR_2} and the present work is the first measurement of this decay employing the total absorption $\gamma$-ray spectroscopy technique. 

$^{100}$Mo has been used along with other isotopes for the 0$\nu$ double $\beta$-decay search in the NEMO 3 experiment, mentioned already. This experiment uses a tracking device and a calorimeter to measure different samples of double $\beta$-decay isotopes. Apart from NEMO 3, there are two experiments based on $^{100}$Mo to search for the 0$\nu$ decay. One is AMoRE (Advanced Mo based Rare process Experiment) \cite{AMORE} based on $^{40}$Ca$^{100}$MoO$_4$ scintillator crystals. The other is MOON (Mo/Majorana Observatory Of Neutrinos) \cite{MOON}, that uses a $^{100}$Mo sheet inserted between two NaI(Tl) detectors. Both experiments exploit the reasonable cost of enrichment in $^{100}$Mo and the large $Q_{\beta\beta}$ that make this isotope attractive for double $\beta$-decay studies.

The decay of $^{100}$Tc $\rightarrow$ $^{100}$Ru has also recently attracted attention in the framework of a different neutrino related topic \cite{Huber2016}. This decay has been identified as an important contributor to a new type of flux-dependent correction to the antineutrino spectrum produced in nuclear reactors. This correction takes into account the contribution of the $\beta$-decay of nuclides that are produced by neutron capture of long-lived fission products. The correction is non-linear in neutron flux, because the process depends on a fission process to produce the fission product ($^{99}$Tc) followed by a neutron capture. For that reason, a better knowledge of the individual $\beta$ branches of this decay can also contribute to a better determination of this correction, of interest for neutrino-oscillation experiments. 

The study of this decay is also of interest from the point of view of nuclear structure. $^{100}$Tc lies in a region of the nuclear chart, where shape effects and shape transitions could play an important role in the evolution of the nuclear structure \cite{Moller_PRL_Axial} and hence in $\beta$-decay rates. The total absorption technique has been used to study shape effects in the parent nucleus, based on the measured $B$(GT) strength distribution in the daughter \cite{KikeShape, PRC_Poirier,  PRC_AnaBelen, PRC_Esther, PRC_Briz}.

\subsection{Total Absorption Spectroscopy}

As already mentioned in the introduction, the $\gamma$-rays emitted in the decay of $^{100}$Tc $\rightarrow$ $^{100}$Ru have only been measured with HPGe detectors. In such conventional high-resolution experiments, $\beta$ intensity to states at high excitation in the daughter nucleus may remain undetected due to the relatively poor efficiency of the HPGe detectors used. This experimental problem, the so-called \textit{Pandemonium} effect \citep{Pandemonium}, can be avoided with the Total Absorption $\gamma$-ray Spectroscopy (TAGS) technique. Experiments performed in the past at GSI, ISOLDE and Jyv\"askyl\"a using this technique have confirmed its potential \cite{Dy,PRC_Rare,LoliTc,PRC_Briz,KikeShape,PRC_Esther,PRC_AnaBelen,PRC_Poirier,DecayHeat,vTAS_PRL}. Moreover, methods were developed by the Valencia group to extract precise $\beta$ intensities from the data \cite{TAS_algorithms,TAS_MC,TAS_pileup,TAS_level}. Looking for possible weak branches that remained undetected in high-resolution studies is the reason why we considered measuring the $\beta$-decay of $^{100}$Tc $\rightarrow$ $^{100}$Ru with the TAGS technique. This could improve the experimental constraints on nuclear models used in double $\beta$-decay calculations for the A=100 system.

\section{Experiment}

The measurement of the $^{100}$Tc $\rightarrow$ $^{100}$Ru decay was performed at the upgraded IGISOL IV facility of the University of Jyv\"askyl\"a (Finland) \citep{Moore_IGISOLIV} in February 2014. For this experiment, the new Decay Total Absorption $\gamma$-ray Spectrometer (DTAS) \citep{DTAS_design}, made of NaI(Tl) crystals, was used in the eighteen-module configuration \citep{NIMB_DTAS}. The $^{100}$Tc nuclei were produced from a Mo target (97.42$\%$ enrichment of $^{100}$Mo) by means of a (p,n) reaction with protons of 10~MeV from the MCC30 cyclotron that were slowed down to 8~MeV with a degrader to maximize the reaction yield.

\begin{figure}[h]
\includegraphics[width=0.4 \textwidth]{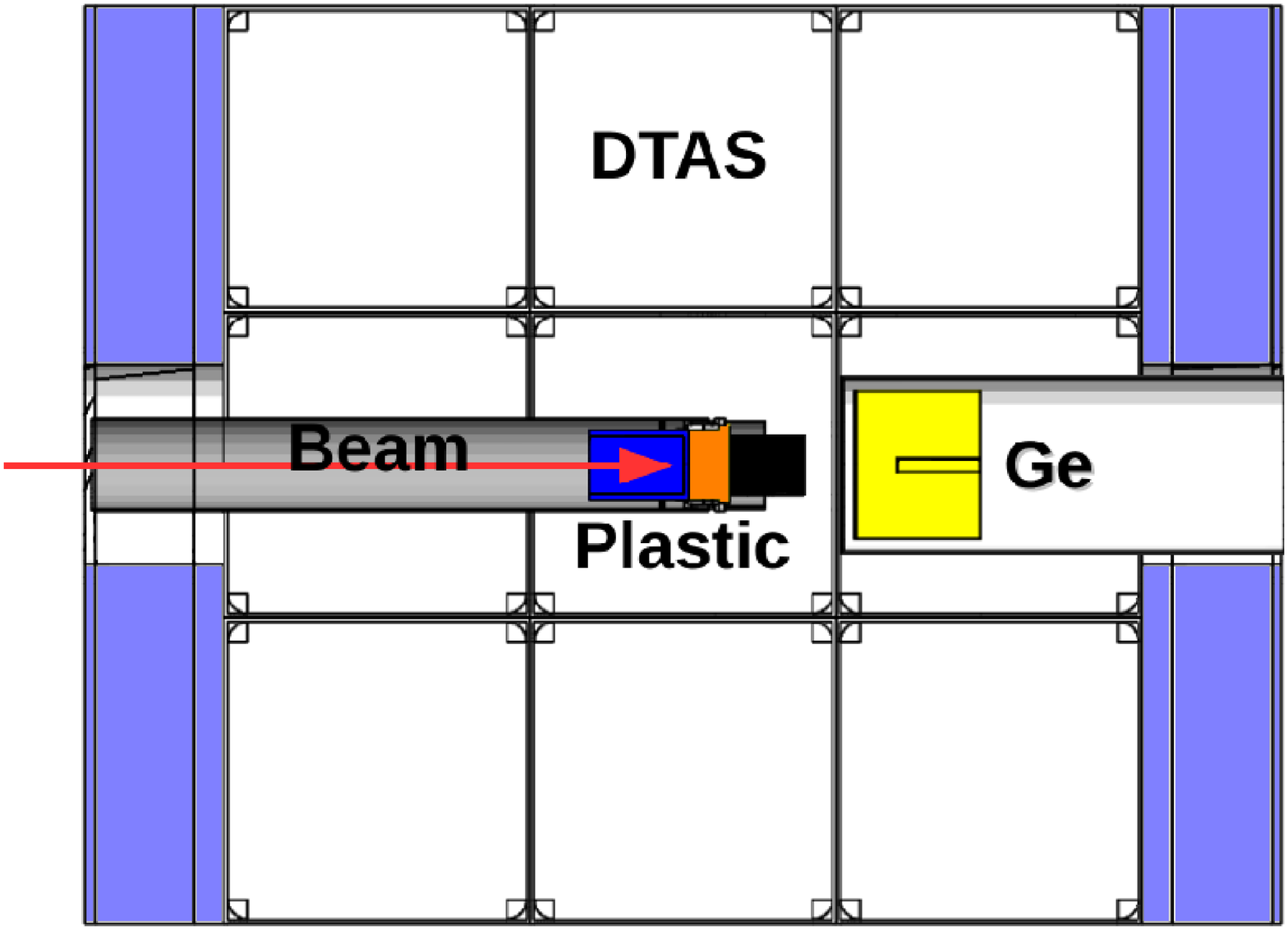} 
\includegraphics[width=0.5 \textwidth]{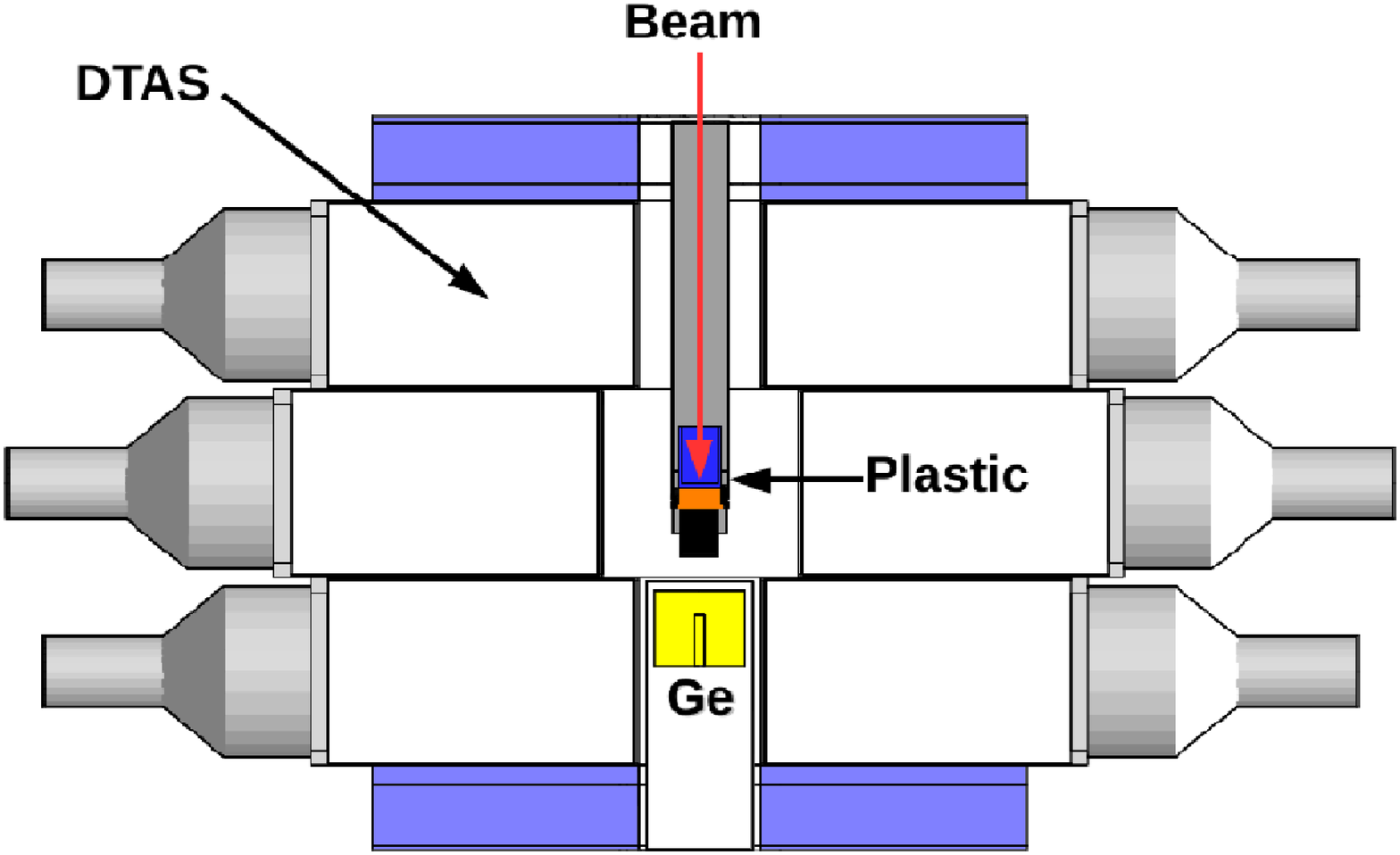}
\caption{\label{Setup100Tc} Experimental set-up for the measurement of the decay of $^{100}$Tc. A lateral cut (up) and a horizontal cut (bottom) are shown. The DTAS detector (in white) surrounded by the shielding (in violet), the beam pipe (in grey), the $\beta$ plastic detector with vase-shaped geometry (in blue) and the HPGe detector (in yellow) are depicted.}
\end{figure}

Since $^{100}$Tc decays to a stable daughter nucleus, there was no need to remove activity after implantation to eliminate the contamination from the descendants. Accordingly, after the purification in the JYFLTRAP double Penning trap \cite{JYFLTRAP} the activity was implanted directly at the bottom of a plastic $\beta$-detector, which has a vase-shaped geometry \cite{NIMA_Vase}. This detector was placed at the centre of the DTAS detector system. The DTAS spectrometer was surrounded by shielding composed of stainless steel sheets, lead bricks and aluminium, which served to reduce the background counting rate by one order-of-magnitude. The set-up was completed with a HPGe detector placed behind the $\beta$ plastic detector, as shown in the schematic view of Figure \ref{Setup100Tc}.

\begin{figure}[h]
\includegraphics[width=0.5 \textwidth]{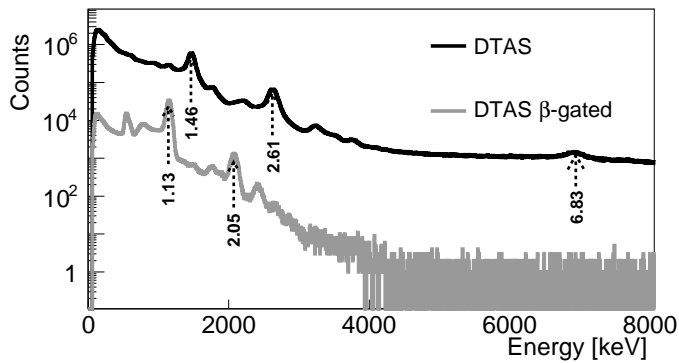}
\caption{\label{100Tc_exp} Experimental measurement of the decay of $^{100}$Tc with DTAS. The spectrum without any condition is shown in black, while the spectrum in coincidence with the $\beta$ plastic detector ($\beta$-gated) is presented in grey. The energies of some of the most relevant peaks are shown in MeV.}
\end{figure}

The total absorption signal of the DTAS detector was reconstructed offline by summing the signals from the eighteen individual modules and applying a method to correct possible changes in the photomultiplier gain based on an external reference detector, as described in \cite{NIMA_DTAS}. In our analog-to-digital converter (ADC) the spectrum of each individual module covered a range of 15~MeV with a threshold of $\sim$ 90~keV. The resulting software sum for the total 21~hours of measuring time is presented as the black line in Figure \ref{100Tc_exp}, and it is dominated by the background. In particular, the characteristic peaks at 1460.8~keV ($^{40}$K) and at 2614.5~keV ($^{208}$Tl) can be clearly seen, as well as the neutron capture peak in the $^{127}$I of the NaI(Tl) crystals at around 6.83~MeV.

\begin{figure}[h]
\includegraphics[width=0.5 \textwidth]{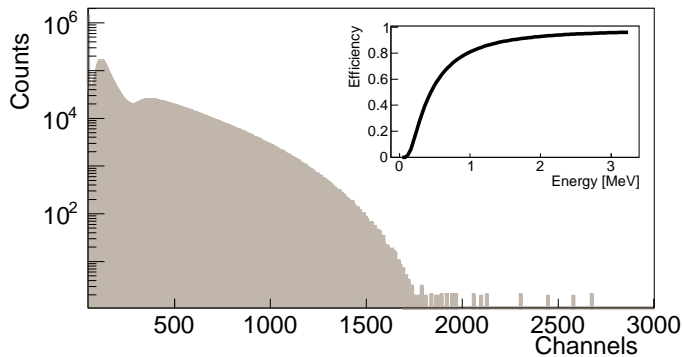}
\caption{\label{Beta} Experimental $\beta$ spectrum for the $^{100}$Tc decay measured with a vase-shaped plastic scintillator. The simulated efficiency curve of the detector is shown in the top-right inset.}
\end{figure}

In order to clean the spectrum and select only those events coming from the $\beta$-decay, coincidences with $\beta$ particles were required, as shown by the grey line in Figure \ref{100Tc_exp}. For this purpose we used a vase-shaped plastic detector \cite{NIMA_Vase} of 35~mm external diameter and 50~mm length with 3~mm thickness in the lateral walls and in the bottom. The plastic detector was covered internally by a thin aluminized-mylar reflector in order to improve the light collection. The $\beta$ spectrum and the efficiency curve calculated with Monte Carlo (MC) simulations are shown in Figure \ref{Beta}. Due to the geometry of this detector, a careful characterization with MC simulations using optical photons was needed in order to understand and reproduce the shape of the resulting $\beta$ spectra and calculate accurately the dependence of the $\beta$-efficiency as a function of the $\beta$-end point energy \cite{NIMA_Vase}. The amount of light collected from the lateral walls was shown to be smaller than the light collected from the bottom, thus producing the bump at the beginning of the spectrum shown in Figure \ref{Beta}. The higher part of the light distribution above this bump is essentially due to interactions in the bottom of the detector. Two different energy thresholds were identified depending on the point where energy is deposited in the detector: 30~keV for the bottom and 100~keV for the lateral walls (see reference \cite{NIMA_Vase} for more details).   

\section{Analysis and results}

The analysis was performed with the experimental $\beta$-gated spectrum shown in Figure \ref{100Tc_exp}. The $\beta$-intensity distribution was obtained with a de-convolution method using the spectrometer response to the decay \cite{TAS_algorithms}, to solve the inverse problem represented by:

\begin{center}
\begin{equation}\label{inverse}
d_i=\sum\limits_{j}R_{ij}(B)f_j+C_i
\end{equation}
\end{center}

\noindent where $d_i$ is the number of counts in channel $i$ of the spectrum, $f_j$ is the number of events that fed level $j$ in the daughter nucleus, $C_i$ is the contribution of all contaminants to channel $i$, and  $R_{ij}(B)$ is the response function of the detector that represents the probability that feeding to the level $j$ gives a count in channel $i$ of the experimental spectrum. This response function is calculated by means of MC simulations, and it is unique for each detector and each decay scheme \cite{TAS_algorithms}. In particular, it depends on the de-exciting branching ratio matrix $B$ of the levels in the daughter nucleus. The calculation of the branching ratio matrix is based partially on the known decay information for the levels at low excitation, that is taken from the literature, assuming that they are well known from high-resolution measurements. According to the Reference Input Parameter Library (RIPL-3) \cite{RIPL-3}, the level scheme of $^{100}$Ru is complete up to a level at 3.072~MeV. Accordingly our first choice for the known level scheme includes all levels up to this level. A second choice was to consider all levels up to the level at 2.387~MeV, the last level with a known spin-parity assignment seen in $\beta$-decay \cite{100Tc_HR_1,100Tc_HR_2}. From the last known level included up to $Q_{\beta}$=3.204~MeV, a continuum region with 40~keV bins is defined with branching ratios based on the statistical model \cite{TAS_level}. This complements the decay scheme in the energy window of the $\beta$-decay. All parameters used for the statistical model calculation are extracted from RIPL-3 \cite{RIPL-3} and summarized in Table \ref{parameters}, with Photon Strength Function (PSF) and deformation parameters based on \cite{PSF} and \cite{DeformationPar_exp}, respectively. The level density parameter ``a'' at the neutron binding energy is obtained from Enhanced Generalized Superfluid Model (EGSM) calculations. The Hartree-Fock-Bogoliubov (HFB) plus combinatorial nuclear level density \cite{Gorieli1,Gorieli2} has been used, with C and P correction parameters of 0.01596 and 0.33071, respectively. 

\begin{table*}[b]
\begin{ruledtabular}
\caption{\label{parameters} Parameters used in the statistical model calculation of the branching ratio matrix (B) of the daughter nucleus $^{100}$Ru.}
  \begin{tabular}{@{}C{2.5cm}C{2cm}C{1cm}C{1cm}C{1cm}C{1cm}C{1cm}C{1cm}C{1cm}C{1cm}C{1cm}@{}}
    Level-density parameter
    & Deformation parameter
    & \multicolumn{9}{c}{Photon strength function parameters}
    \\ \cmidrule(lr){3-11}
    &  & \multicolumn{3}{c}{E1} & \multicolumn{3}{c}{M1} & \multicolumn{3}{c}{E2}   \\
    \cmidrule(r){1-1}\cmidrule(lr){2-2}\cmidrule(lr){3-5}\cmidrule(lr){6-8}\cmidrule(l){9-11}
    a & $\beta$  & E & $\Gamma$ & $\sigma$ & E & $\Gamma$ & $\sigma$ & E & $\Gamma$ & $\sigma$  \\
    &  & [MeV] & [MeV] & [mb] & [MeV] & [MeV] & [mb] & [MeV] & [MeV] & [mb]  \\
    \\ \cmidrule(lr){1-11}
    8.4341 &  0.2148
    & \begin{tabular}{@{}c@{}}14.531\\ 17.416\end{tabular} 
    & \begin{tabular}{@{}c@{}}4.201\\ 5.926\end{tabular} 
    & \begin{tabular}{@{}c@{}}78.421\\ 111.167\end{tabular} 
    & 8.847 & 4.000 & 2.277 & 13.594 & 4.910 & 2.358  \\  
  \end{tabular}
\end{ruledtabular}
\end{table*}

Once the branching ratio matrix is constructed, the response function $R_{ij}(B)$ can be calculated recursively from mono-energetic $\gamma$-ray MC responses, folded with the response to the $\beta$ continuum for each level \cite{TAS_MC}. For the simulations we use the Geant4 package \cite{GEANT4}, using a detailed description of the geometry of the set-up (the DTAS spectrometer, the ancillary detectors and the beam pipe). Moreover, the MC simulations include the non-proportionality of the light yield in NaI(Tl) in the form described in \cite{TAS_MC}. The inclusion of this process has been shown to be crucial \cite{TAS_MC,NIMA_DTAS} in the analysis of TAS data obtained with spectrometers made of this material. The Geant4 MC simulations were validated for this geometry by comparison with measurements of well-known radioactive sources ($^{24}$Na, $^{60}$Co, $^{137}$Cs, $^{22}$Na, and $^{152}$Eu-$^{133}$Ba) \cite{NIMA_DTAS}.

In addition, we investigated the sensitivity of the $\beta$-detector to $\gamma$-rays. This can introduce distortions in the $\beta$-gated TAGS spectrum. For a realistic estimation, we made a MC simulation with decay cascades generated with the DECAYGEN event generator \cite{TAS_level}. As input to this event generator we use the branching ratio matrix and the $\beta$ intensity distribution from our analysis. The output consists of an event file where the primaries are labelled and can be identified. By simulating the $\beta$-particles and $\gamma$-rays from the event file, and comparing with a simulation  with only $\beta$-particles, we deduced that around $0.2\%$ of the counts in the total simulation for the $^{100}$Tc are coming from the interaction of $\gamma$-rays with our plastic detector, which represents a negligible distortion.

\begin{figure}[h]
\includegraphics[width=0.5 \textwidth]{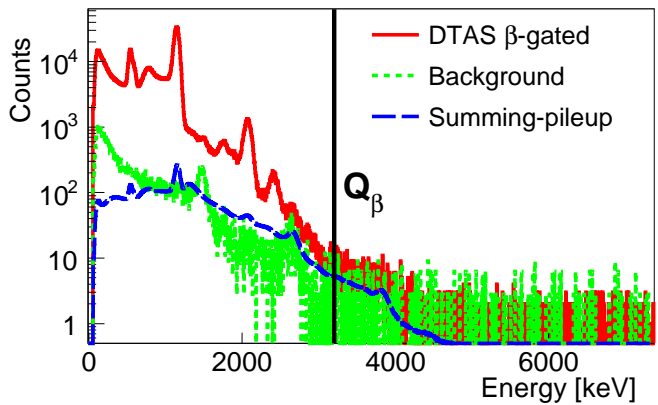}
\caption{\label{100Tc_contaminants} Contaminants of the $\beta$-gated experimental spectrum (red line) in the decay of $^{100}$Tc: background (dotted green) and summing-pileup (dashed blue).}
\end{figure}

In a TAGS analysis it is crucial to identify all the sources of contamination. In this case, although we are considering coincidences with the $\beta$ detector, the large ground state feeding intensity of this decay, which is around 90$\%$ as we shall see later, together with the high efficiency of the TAS gives rise to a non-negligible number of random coincidences of the $\beta$-particles with the environmental background in DTAS. This dominates the DTAS spectrum if no coincidence conditions are imposed, as shown in Figure \ref{100Tc_exp}. The contribution of this contamination was obtained using the two main peaks at 1460.8~keV and at 2614.5~keV that were mentioned earlier. Apart from the environmental background, we have to consider the contribution of the summing-pileup of signals. To deal with this we follow the procedure explained in \cite{NIMA_DTAS} that has already been applied successfully in previous works \cite{vTAS_PRL,Zak_PRL,vTAS_PRC}. It is based on the random superposition of two stored events within the ADC gate length. This contribution is normalized with a theoretical expression based on \cite{TAS_pileup}. In Figure \ref{100Tc_contaminants} the contribution of the contaminants is shown together with the $\beta$-gated spectrum. 

\begin{figure}[h]
\includegraphics[width=0.5 \textwidth]{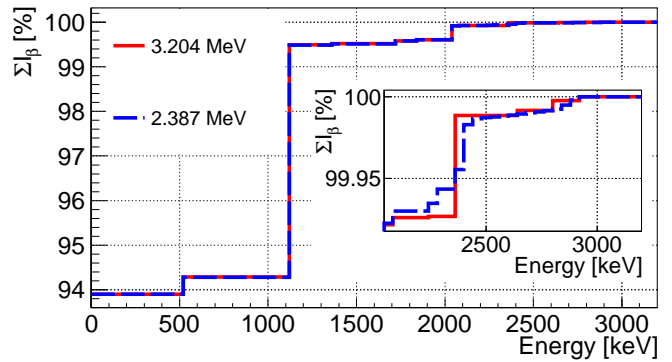}
\caption{\label{100Tc_cuts} $\beta$-intensity distribution extracted from the TAGS analysis with a known level scheme up to 3.072~MeV compared to the distribution with a known level scheme up to 2.387~MeV. A zoom in the last MeV of the $Q_{\beta}$ window in $^{100}$Ru is presented in the inset to show the differences between the two analysis.}
\end{figure}

\begin{figure}[h]
\includegraphics[width=0.5 \textwidth]{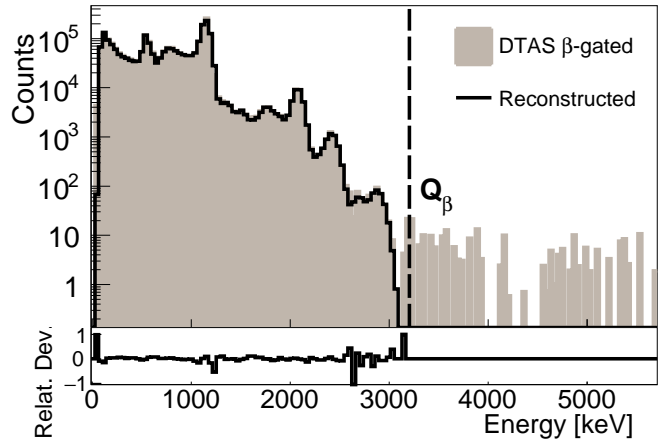}
\caption{\label{100Tc_analysis} Result of the analysis of the $^{100}$Tc decay: $\beta$-gated experimental spectrum after subtracting the contaminants (filled grey) is compared with the reconstructed spectrum after the analysis (black). The reconstructed spectrum is obtained by convoluting the response function with the final accepted feeding distribution.}
\end{figure}

The analysis was carried out by applying the expectation maximization (EM) algorithm to extract the $\beta$-feeding distribution \cite{TAS_algorithms}. The quality of the analysis can be checked by the comparison of the experimental spectrum with the spectrum reconstructed with the $\beta$-intensities obtained in the analysis convoluted with the response function of the spectrometer by using Equation \ref{inverse}. We have observed that there is no noticeable difference if we consider the known level scheme up to 3.072~MeV or up to 2.387~MeV, as can be seen in Figure \ref{100Tc_cuts}, where both $\beta$-intensity distributions are compared. Moreover, from the analysis it was concluded that allowing feeding only to states observed in the high-resolution measurement of the $\beta$-decay was enough to obtain a good reproduction of the spectrum. However, the fit at high energies in the analysis with the known part up to 3.072~MeV is improved if we consider an additional 2$^+$ level at 2.934~MeV that was not seen in previous $\beta^-$ decay studies, but was seen in electron capture studies from $^{100}$Rh \cite{ENSDF100}. The improvement of the fit with $\beta$ intensity at this energy is also seen in the analysis performed with the known level scheme up to 2.387~MeV, where the last level populated in the continuum is at 2.940~MeV. The level at 2.934~MeV, according to the information from the Evaluated Nuclear Structure Data File (ENSDF), de-excites with a single $\gamma$-ray of 2934 keV to the ground state and has spin-parity values of 1$^+$ or 2$^+$. In the RIPL-3 database \cite{RIPL-3}, a spin-parity assignment of 2$^+$ is suggested and this was our assumption in the analysis. However, we also tested the 1$^+$ spin-parity assignment as a possibility for this level, with very similar results. The final $\beta$ intensity distribution was obtained with the known level scheme up to 3.072~MeV, and it is presented in the fourth column of Table \ref{tableIb}. The quality of the final analysis is shown in Figure \ref{100Tc_analysis}.

\begin{table*}[b]
\caption{\label{tableIb}$I_{\beta}$ and $\log ft$ values obtained with DTAS compared with the information from ENSDF \cite{ENSDF100}. The theoretical calculated values are also listed, and they have been computed by using the 'linear' model by adopting the value $g_A$=0.40. The experimental $0^+$ state at 1.741~MeV is missing, since it is likely to be a three-phonon state in terms of structure, and such states are outside the model space of the theoretical framework used for the computations (details in Section \ref{Theory}).}
\begin{ruledtabular}
\begin{tabular}{ccccccc}
Energy [MeV] & $J^P$ & $I_{\beta}$ ENSDF [$\%$] & $I_{\beta}$ DTAS [$\%$] & $\log ft$ ENSDF & $\log ft$ DTAS  & $\log ft$ theory\\ \hline
0.000 &
$0^+$ &
93.3(1\footnote{for further discussion see text.}) & 93.9(5) &
4.591(6)\footnote{\label{note1} This $\log ft$ value has been calculated with the $\log ft$ program of the National Nuclear Data Center (NNDC) \cite{logftNNDC} that uses ENSDF evaluated data as input (the $\beta$ intensity from the third column). It differs slightly from the $\log ft$ value of the evaluation \cite{ENSDF100}.} &
4.588(6) &
4.63
\\
0.538 &
$2^+$ &
0.75(14) & 0.39(5) & 
6.35(9)$^{\text{\ref{note1}}}$ &
6.63(6) &
5.88
 \\
1.130 &
$0^+$ &
5.36(13) & 5.20(40) &
5.04(1) & 
5.05(4)  &
6.06
\\
1.362 &
$2^+$ &
0.030(4) & 0.026(8) &
7.1(1) &
7.15(14) &
7.35
 \\
1.741 &
$0^+$ &
0.066(3) & 0.062(6) & 
6.34(2) &
6.37(5) &
-
 \\
1.865 &
$2^+$ &
0.030(4) & 0.029(3) & 
6.54(6) &
6.55(5) &
-
 \\
2.052 &
$0^+$ &
0.36(5) & 0.31(2) &
5.21(6) &
5.27(3)  &
5.30
\\
2.099 &
$2^+$ &
0.0073(7) & 0.0045(40) &
6.83(5) &
7.04(40)  &
-
\\
2.241 &
$2^+$ &
0.0013(7) & 0.0006(5) &
7.36(20)$^{\text{\ref{note1}}}$ &
7.69(80)  &
-
\\
2.387 &
$0^+$ &
0.063(4) & 0.062(6) &
5.41(3) &
5.42(5)  &
5.27
\\
2.660 &
$2^+$ &
0.0046(10) & 0.0032(30) &
5.9(1) &
6.1(10)  &
6.24
\\
2.838 &
$2^+$ &
0.006(3) & 0.006(1) &
5.2(2) &
5.22(8)  &
5.73
\\
2.934 &
$2^+$ &
- & 0.0024(9) & 
- &
5.18(20) &
5.64
 \\
\end{tabular}
\end{ruledtabular}
\end{table*}

For the evaluation of the uncertainties in the $\beta$-intensities resulting from the analysis and presented in Table \ref{tableIb}, several sources of systematic error were considered (statistical errors are negligible in comparison). First, the normalization factors of the contaminants were varied and the impact on the $\beta$-intensities evaluated. We have found that the reproduction of the experimental spectrum allows a change of up to $\pm$50$\%$ for the normalization factor of the background, and $\pm$10$\%$ for the summing-pileup. The impact of the effect of the $\beta$ detector efficiency has also been studied by changing the threshold value in the MC simulation by $\pm$30$\%$. Finally, the Maximum Entropy (ME) algorithm \cite{TAS_algorithms} has been applied instead of the EM algorithm in order to check the influence of the method of de-convolution. By combining all of these sources of uncertainty, we have estimated the possible systematic errors in the analysis. 

\begin{figure}[h]
\includegraphics[width=0.5 \textwidth]{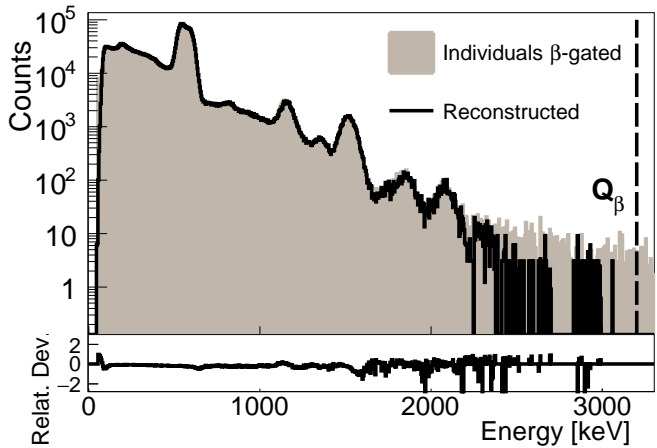}
\caption{\label{100Tc_individuals} Comparison of the 18 individual experimental spectra summed without contaminants (filled grey) with the reconstructed spectrum after the analysis (black).}
\end{figure}

Finally, as a crosscheck of the consistency of the analysis, we have also calculated the $I_{\gamma}$ values de-exciting the main levels populated in the decay, using our branching ratio matrix and our $I_{\beta}$ distribution. The result is presented in Table \ref{tableIg} and it shows a nice agreement with the high-resolution $\gamma$ intensities. Furthermore, the segmentation of the spectrometer allows us to check the reproduction of the individual-module spectra of DTAS. A simulation using the DECAYGEN event generator \cite{TAS_level} with the branching ratio matrix and the $\beta$ intensity distribution from our analysis as input, reproduces nicely the sum of the 18 single-crystal spectra when compared with experiment, as shown in Figure \ref{100Tc_individuals}. 

\begin{table}[b]
\caption{\label{tableIg}Main values of absolute $\gamma$ intensities de-exciting the levels in $^{100}$Ru per 100 decays.}
\begin{ruledtabular}
\begin{tabular}{ccc}
Energy [keV] & $I_{\gamma}$ ENSDF & $I_{\gamma}$ DTAS  \\ \hline
539.48  &  0.066 & 0.060 \\
1130.25  &  0.054 & 0.052 \\
1362.21  &  0.001 & 0.001 \\
1740.95  &  0.001 & 0.001 \\
2051.51  &  0.004 & 0.003 \\
2387.12  &  0.001 & 0.001 \\
\end{tabular}
\end{ruledtabular}
\end{table}

\begin{figure}[h]
\includegraphics[width=0.5 \textwidth]{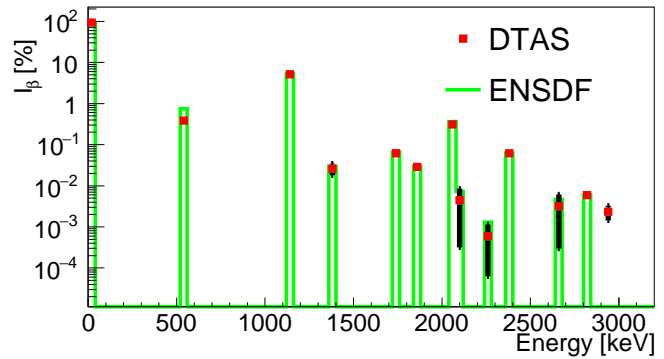}
\caption{\label{100Tc_feedings} $\beta$-intensities of the $^{100}$Tc decay from ENSDF (green) and from the TAGS analysis (red).}
\end{figure}

\begin{figure}[h]
\includegraphics[width=0.5 \textwidth]{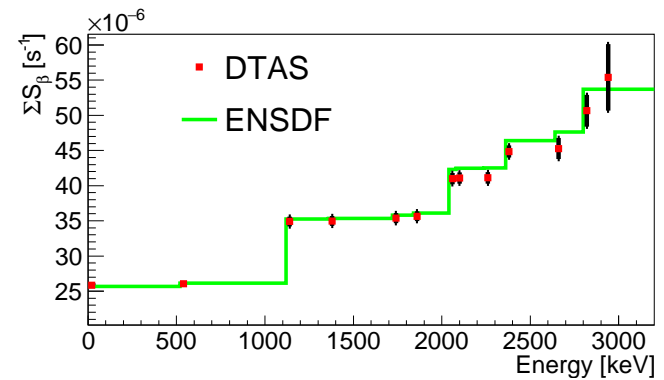}
\caption{\label{100Tc_strength} Comparison of the accumulated $\beta$-strength of the $^{100}$Tc decay for the data from ENSDF (green) and the data obtained with DTAS (red).}
\end{figure}

The information for the $I_{\beta}$ from ENSDF \cite{ENSDF100} is compared with the result of the analysis in Figure \ref{100Tc_feedings}. In Table \ref{tableIb} we present the $I_{\beta}$ values and $\log ft$ values corresponding to this comparison. The accumulated strength is also calculated in both cases and compared in Figure \ref{100Tc_strength}. 

From the data compiled in Tables \ref{tableIb} and \ref{tableIg}, it can be concluded that the TAGS results confirm the high-resolution results in this case. All $\beta$ intensities are in agreement within the errors, except for the first 2$^+$ state. In general, $\beta$ intensities obtained exhibit relative differences of $<15\%$ with respect to ENSDF. The new TAGS data give a slightly larger ground state feeding intensity ($<1\%$ difference) and a population of the first 2$^+$ state $\sim50\%$ smaller. Intensities to levels at 2.099~MeV and 2.241~MeV have relative differences with ENSDF of $\sim40\%$ and $\sim50\%$ respectively. These levels are weakly populated, and the $\beta$ intensities are determined with large errors in our analysis. Similarly, the intensity to the level at 2.660~MeV also has a large error because it is strongly affected by the 2614.5~keV peak in the environmental background, and the intensity differs by $\sim30\%$ from the evaluated value. 

Concerning the most important branch of the decay, the ground state feeding, we obtained a value of 93.9(5)$\%$, in comparison with the 93.3(1)$\%$ value from ENSDF \cite{ENSDF100}. However, it should be noted that the quoted value in ENSDF has a quite small error. In the only high-resolution reference that gives absolute $\gamma$-intensities with errors \cite{100Tc_HR_1}, the $\gamma$-ray with 100$\%$ relative intensity (539.6~keV) is measured with an absolute intensity of 6.6(6)$\%$. Using this number we have evaluated the ground state feeding intensity, and obtained 93.3(6)$\%$, where the error is calculated by applying the conventional method for error propagation. If we consider the updated error given by the same authors in a subsequent publication \cite{100Tc_HR_1_last}, with 6.6(5)$\%$, a ground state feeding intensity of 93.3(4)$\%$ is obtained. In both cases, the error is larger than the ENSDF value, and our value of 93.9(5)$\%$ is in reasonable agreement with them. It is worth mentioning that in the EC decay study from \cite{EC100}, a 6.6(3)$\%$ absolute intensity is obtained for the 539.6~keV $\gamma$-ray. An evaluation of the ground state feeding intensity by combining this error and the relative intensities from \cite{100Tc_HR_1} gives a 93.3(2)$\%$. Furthermore, we have also calculated the ground state feeding by applying a $\beta$-$\gamma$ counting method for TAGS data proposed by Greenwood et al. \cite{Greenwood_GS}. Our preliminary calculation with this method gives a value of 92.8(5)$\%$, closer to the value from high-resolution measurements. However, we should note that with this ground state feeding intensity we do not obtain as good reproduction of the low energy part of the TAGS spectrum as with the value reported in Table \ref{tableIb}.

\section{Theoretical description of the results}\label{Theory}

As mentioned in the introduction, the original goal of this work was to contribute with an independent 
measurement of the $^{100}$Tc $\rightarrow$ $^{100}$Ru $\beta$-decay to the overall knowledge of the $A=100$ system and thus provide a better experimental constraint on the nuclear models used in double $\beta$-decay calculations. In this section, we will give a few details of the calculations performed using the  quasiparticle random-phase approximation (QRPA) for this decay, and we will compare the results of the TAGS analysis with these calculations. 

\subsection{Description of the nuclear model}

The wave functions of the nuclear states involved in the 
$\beta$-decay transitions of $^{100}$Tc into $^{100}$Ru are calculated in this case using QRPA in a realistically large 
single-particle model space spanned by the single-particle orbitals 
1p-0f-2s-1d-0g-0h for both protons and neutrons, with all 
spin-orbit partners included. The calculated $2^+$ states, except for $2^+_2$, and $0^+$ states, except for $0^+_1$, in $^{100}$Ru 
are assumed to be basic excitations (one-phonon states) of the 
charge-conserving QRPA (ccQRPA) \cite{Suhonen2007}, whereas
the $0^+_1$ and $2^+_2$ states are assumed to consist of two $2^+_1$ ccQRPA
phonons as discussed in \cite{Civitarese1994}. 
The $J^+$ ground state of the nucleus $^{100}$Tc is generated by the usual
proton-neutron QRPA (pnQRPA) \cite{Suhonen2007}.
The one- and two-phonon states in $^{100}$Ru are then connected to 
the $1^+$ ground state of $^{100}$Tc by transition amplitudes obtained 
from a higher-QRPA framework called the multiple-commutator model 
(MCM), first introduced in \cite{Suhonen1993} and further extended 
in \cite{Civitarese1994}. The MCM framework has been used on many occasions in
the past in $\beta$-decay and double-$\beta$-decay calculations, as described in
\cite{Suhonen2012}.

For the Gamow-Teller $\beta^-$ transitions $1^+\to 0^+,2^+$ we can define \cite{Suhonen2007}:

\begin{eqnarray}
\nonumber
\log ft = \log_{10} (f_0t_{1/2}[\textrm{s}]) = 
\log_{10}\left[\frac{6147}{B_{\rm GT}}\right]\ , \\
B_{\rm GT}=\frac{g_{\rm A}^2}{3}\vert\mathcal{M}_{\rm GT}\vert^2 \ ,
\end{eqnarray}
where $f_0$ is a phase-space factor, $t_{1/2}$[s] is the partial half-life of a $\beta$ 
transition in seconds and $g_{\rm A}$ is the weak axial-vector coupling constant
with its bare one-nucleon value $g_{\rm A}=1.27$. The quantity $\mathcal{M}_{\rm GT}$ is 
the Gamow-Teller transition matrix element to be computed by the MCM method.

The single-particle energies were first generated by the use of a 
spherical Coulomb-corrected Woods-Saxon (WS) potential, with the
global parametrization of Ref.~\cite{Bohr1969}. The BCS approximation was used to 
define the quasiparticles needed for the pnQRPA calculations of the wave functions in the 
nucleus $^{100}$Tc and the ccQRPA calculations of the wave functions in the final 
nucleus $^{100}$Ru. The Bonn-A G-matrix \cite{Holinde1981} has been used as the starting 
point for the two-body interaction and it has been scaled separately for the pairing and
proton-neutron multipole channels \cite{Suhonen1988_1,Suhonen1988_2}. The 
pairing matrix elements are scaled by a common factor, separately for protons and 
neutrons, and in practice these factors are fitted so that the lowest quasiparticle 
energies obtained from the BCS match the experimental 
pairing gaps for protons and neutrons respectively. 

The particle-hole and particle-particle parts of the proton-neutron two-body interaction
in the pnQRPA calculation are scaled by the particle-hole parameter $g_{\rm ph}$ and 
particle-particle parameter $g_{\rm pp}$, respectively \cite{Suhonen1988_1,Suhonen1988_2}. The value of the 
particle-hole parameter was fixed by the available 
systematics \cite{Suhonen2007} on the location of the Gamow-Teller 
giant resonance (GTGR) for $1^+$ states. The value of $g_{\rm pp}$ is not fixed a priori
and it is a free parameter in the model. Its value regulates the $\beta^-$ decay amplitude 
of the first $1^+$ state in an odd-odd nucleus \cite{Suhonen2005}, as here in the case of the ground state of $^{100}$Tc.
Also the value of the axial vector coupling
constant $g_{\rm A}$ is not known in finite nuclei.
The effective (quenched) value of $g_{\rm A}$ has attracted a lot of attention recently
due to the fact that it plays a crucial role in predictions of the rates of
double $\beta$-decays, which depend on $g_{\rm A}$ to the fourth power. Typically, in the
shell-model calculations in the sd and pf shells a moderate quenching, $g_{\rm A}\sim 1$, 
has been adopted \cite{Wildenthal1983,Martinez-Pinedo1996}. However, a
strong quenching of $g_{\rm A}\sim 0.6$ was reported in the shell-model 
calculations in the mass $A=90-97$ region in Ref.~\cite{Juodagalvis2005}.
In a more recent shell-model study \cite{Caurier2012} values of about 
$g_{\rm A}\sim 0.7$ were obtained in the mass region $A=128-130$ and 
an even stronger quenching of $g_{\rm A}= 0.56$ for $A=136$. Strong quenchings
for $g_{\rm A}$ have also been obtained in the framework of the pnQRPA 
\cite{Faessler2008,Suhonen2013,Suhonen2014,Pirinen2015} and in the
interacting boson approximation calculations \cite{Barea2013,Yoshida2013,Barea2015}.
A combined global analysis of the values of $g_{\rm pp}$ and $g_{\rm A}$ was performed 
in the pnQRPA approach in \cite{Pirinen2015}. The measured Gamow-Teller
ground-state-to-ground-state $\beta$-decay rates were compared with the computed ones
within the mass range $A=100-136$. In the present calculations we adopt the values
$g_{\rm pp}=0.70$ and $g_{\rm A}=0.40$ directly from this global analysis and use them
to compute the $\log ft$ values for all $\beta^-$ decay transitions in this work.

For the ccQRPA the $g_{\rm pp}$ parameter was kept in the default, pure G-matrix value
$g_{\rm pp}=1.00$ and $g_{\rm ph}$ was fixed to reproduce the experimental 
excitation energy $E(2^+_1)=538\,\textrm{keV}$
of the $2^+_1$ state in $^{100}$Ru by the ccQRPA calculations. The $0^+_1$, $2^+_2$ and
$4^+_1$ states are assumed to belong to a two-phonon triplet where the degeneracy of 
the states is lifted by their interactions with the one-phonon states, as discussed in
\cite{Delion2003}. In the MCM description non-interacting
two-phonon states are used at exactly twice the energy of the $2^+_1$
state, and no mixing with the one-phonon states is assumed. Hence, in
the present MCM calculations the states $0^+_1$, $2^+_2$ and $4^+_1$
share the common energy of $1.076\,\textrm{MeV}$.

\subsection{Discussion}

The resulting $\log ft$ values obtained from this calculation are presented together with the 
experimental ones in Table \ref{tableIb}. As one can see, the ground-state $\log ft$ value is 
well reproduced due to the features of the global fit of
\cite{Pirinen2015}. However, the $\log ft$ predictions for transitions to the $2^+_1$ and $0^+_1$ 
states fail. This seems to be a characteristic problem with the transitions to excited
states in $^{100}$Ru since similar difficulties were faced in the earlier calculations
of \cite{Griffiths1992,Suhonen1994}. In these studies a 
simultaneous prediction of the two-neutrino double $\beta$-decay rate 
of $^{100}$Mo and the $\beta$-decay rates of $^{100}$Tc
was attempted and the $g_{\rm pp}$ parameter of the pnQRPA was used for this purpose.
By varying $g_{\rm pp}$ and keeping $g_{\rm A}$ moderately quenched ($g_{\rm A}\sim 1$)
a quite good result for the double-$\beta$-decay half-life,
$t_{1/2}^{(2\nu )}=7.66\times 10^{18}$ yr, was obtained when compared
with the present experimental value $t_{1/2}^{(2\nu )}({\rm exp})=(7.1\pm 0.4)\times 10^{18}$ yr
\cite{Barabash2013} with a similar single-particle basis set to that used in the present calculations.
Instead, in the global fit of \cite{Pirinen2015} the same $g_{\rm pp}=0.70$ was used 
for all nuclei within the mass range $A=100-136$ and a half-life three times
longer than the experimental one was obtained by using the linear model
with $g_{\rm A}=0.40$.

One possible obstacle to an accurate theoretical description of the $\beta$-decay properties of 
$^{100}$Tc in the present and earlier calculations is the appearance of deformation effects at 
around mass $A=100$. This is a problem since the pnQRPA calculations conducted here and earlier
are based on a spherical mean field. In an earlier study \cite{Kotila1993} the isotopic
chain $^{98-106}$Ru was studied by using the microscopic anharmonic vibrator approach (MAVA) to
track the possible setting of deformation in the chain. The MAVA uses a realistic nuclear
Hamiltonian to derive equations of motion for the mixing of one- and two-phonon degrees of
freedom starting from the collective phonons of QRPA. This means that the assumption of harmonic vibration in the present calculations is relaxed and the degeneracy of the two-phonon $0^+_1$, $2^+_2$ and $4^+_1$
states, mentioned earlier, is broken by the one-phonon--two-phonon interactions. In the study
\cite{Kotila1993} it was found that the nucleus $^{100}$Ru can be seen as a transitional nucleus
between the anharmonic vibrator $^{98}$Ru and the (quasi-)rotors $^{102-106}$Ru. Furthermore, the theoretical study of \cite{MollerDeform} and the experimental study of \cite{RamanDeform} suggest that $^{100}$Ru possesses a moderate deformation around $0.16-0.21$ implying that $^{100}$Ru is a soft nucleus lying between an anharmonic vibrator and a deformed rotor. For $^{100}$Tc the calculations of \cite{Moller_PRL_Axial} imply a moderate deformation of $0.19$, not far from the deformation of $^{100}$Ru. Hence, $^{100}$Tc can also be considered to be a soft transitional nucleus like $^{100}$Mo. It could be that even this softness, being between a vibrator and a rotor, can affect the $\beta$-decay transitions for $^{100}$Tc $\rightarrow$ $^{100}$Ru in such a way that a perfect description of these $\beta$ transitions becomes impossible with a simple spherical pnQRPA approach.

\section{Conclusions}
In this work we have presented a measurement of the  $^{100}$Tc $\rightarrow$ $^{100}$Ru $\beta$-decay 
using the total absorption $\gamma$-ray spectroscopy technique for the first time. The results of this 
analysis confirm the $\beta$ intensities obtained with HPGe detectors using the high-resolution technique \cite{100Tc_HR_1,100Tc_HR_2}. In particular, the large $\beta$ intensity of the most important branch of the decay, going to the ground state of $^{100}$Ru, has been confirmed. Moreover, a $\beta$-$\gamma$ counting method for TAGS data also gives a ground state feeding intensity in agreement with the TAGS analysis. The remainder of the $\beta$ intensities obtained are also in reasonable agreement with previous results. The largest discrepancies are observed for the first $2^+_2$ state. The best fit in the TAGS analysis is obtained when feeding to a new $2^+_2$ state at 2.934~MeV is introduced. This intensity was not seen in previous $\beta$-decay studies. 

Due to the importance of this decay for double $\beta$ decay studies, it was crucial to confirm with the TAGS technique the available data, avoiding any possible influence of the \textit{Pandemonium} systematic error \cite{Pandemonium}. Although the high-resolution experimental information may look reasonably complete, new intensity was detected in previous TAGS experiments even in apparently well known cases, as in the recent study of the decay of $^{87}$Br \cite{vTAS_PRC}, or in the decay of $^{148}$Dy \cite{Dy}. In addition, this result represents a validation of the good performance of the new experimental set-up formed by the DTAS detector in combination with a vase-shaped plastic detector.     

The decay data have been discussed in the framework of the QRPA calculations, because of their impact 
in double $\beta$-decay calculations. These calculations are in good agreement with TAGS results for the ground state feeding and for the level at 2.052~MeV, with differences in $\beta$ intensity of less than $10\%$. The rest of the calculations lead to $\beta$ intensities differing by between $30\%$ and $70\%$ from TAGS results, except for the $2^+_1$ and $0^+_1$ states in $^{100}$Ru, where discrepancies are a factor of 6 and 10 respectively. These deviations from the measured $\beta$-decay
rates could be due to the small deformation (shape softness) of both the mother and daughter nuclei. Concerning the interesting $2\nu$ double $\beta$-decay there is a slight conflict regarding the adopted effective value of the axial-vector coupling constant $g_{\rm A}$. On the one hand, the $\beta$-decay calculations presented here are performed by adopting the value $g_{\rm A}=0.40$ from the linear model of a global Gamow-Teller $\beta$-decay study. The other model of that study, with constant $g_{\rm A}=0.6$, yields a poorer reproduction of the results for the present decay transitions. On the other hand, the constant $g_{\rm A}=0.6$ model works better for the 2$\nu$ double $\beta$-decay, reproducing almost exactly the $^{100}$Mo 2$\nu$ double $\beta$-decay half-life and many other 2$\nu$ double $\beta$-decay half-lives. In this way the presently discussed $A=100$ triplet -- Mo, Tc and Ru -- continues to be a challenge for nuclear models aiming at a successful description of both the single $\beta$-decays and the $2\nu$ double $\beta$-decay for these nuclei.

\begin{acknowledgments}
AA acknowledges useful discussions with Prof. A Garcia and Dr. S Sjue during the preparation of the experimental proposal. This work has been supported by the Spanish Ministerio de Econom\'ia y Competitividad under Grants No. FPA2011-24553, No. AIC-A-2011-0696, No. FPA2014-52823-C2-1-P, No. FPA2015-65035-P, No. FPI/BES-2014-068222 and the program Severo Ochoa (SEV-2014-0398), by the Spanish Ministerio de Educaci\'on under the FPU12/01527 Grant, by the European Commission under the FP7/EURATOM contract 605203 and the FP7/ENSAR contract 262010, and by the $Junta~para~la~Ampliaci\acute{o}n~de~Estudios$ Programme (CSIC JAE-Doc contract) co-financed by FSE. This work has been partially supported by the Academy of Finland under the Finnish Centre of Excellence Programme 2012-2017 (Project No. 213503, Nuclear and Accelerator Based Programme at JYFL). WG was supported by the UK Science and Technology Facilities Council (STFC) Grant ST/F012012/1 and by the University of Valencia. 
\end{acknowledgments}

\bibliography{100Tc_bib}

\end{document}